\pdfoutput=1
\documentclass[twoside,ngerman,american]{iopart}
\usepackage[T1]{fontenc}
\usepackage[latin9]{inputenc}
\usepackage{color}
\usepackage{babel}
\usepackage{verbatim}
\usepackage{amstext}
\usepackage{amssymb}
\usepackage{setspace}
\usepackage[unicode=true,pdfusetitle,
 bookmarks=true,bookmarksnumbered=false,bookmarksopen=false,
 breaklinks=false,pdfborder={0 0 0},backref=false,colorlinks=true]
 {hyperref}
\hypersetup{
 pdfpagemode=UseNone,pdfstartview=FitH,citecolor=blue,linkcolor=blue}

\makeatletter
\usepackage{iopams}
\usepackage{setstack}

\usepackage{ae}
\usepackage{aecompl}

\usepackage[numbers, sort&compress]{natbib}

\newcommand{\eqref}[1]{(\ref{#1})}

\makeatother

\begin{document}

\title{Demonstration and resolution of the Gibbs paradox of the first kind}

\author{Hjalmar Peters}

\address{Karlsruhe Institute of Technology, Karlsruhe, Germany}

\ead{hjalmar.peters@web.de}
\begin{abstract}
The Gibbs paradox of the first kind (GP1) refers to the false increase
in entropy which, in statistical mechanics, is calculated from the
process of combining two gas systems S1 and S2 consisting of distinguishable
particles. Presented in a somewhat modified form, the GP1 manifests
as a contradiction to the second law of thermodynamics. Contrary to
popular belief, this contradiction affects not only classical but
also quantum statistical mechanics. The present paper resolves the
GP1 by considering two effects: 1. The uncertainty about which particles
are located in S1 and which in S2 contributes to the entropies of
S1 and S2. 2. S1 and S2 are correlated by the fact that if a certain
particle is located in one system, it cannot be located in the other.
As a consequence, the entropy of the total system consisting of S1
and S2 is not the sum of the entropies of S1 and S2.
\end{abstract}


\maketitle

\section{Introduction, motivation, and outline\label{sec:Einleitung}}

In physics, there are two distinct paradoxes which are both known
as the ``Gibbs paradox'' and are often confused with each other.
In the following, the false increase in entropy which, in statistical
mechanics, is calculated from the process of combining two gases of
the same kind consisting of distinguishable particles will be referred
to as the Gibbs paradox of the first kind (GP1) \foreignlanguage{ngerman}{\mbox{\cite{Gibbs1+Gibbs1Reprint1Englisch,SchroedingerEnglisch,HuangEnglisch,Pathria,Peters2010EJP}}.}
The Gibbs paradox of the second kind (GP2) addresses the fact that
the entropy increase, when combining two gases of different kinds,
is independent of the degree of similarity between the two kinds
of gases, and that this entropy increase vanishes discontinuously
at the transition from similarity to sameness \foreignlanguage{ngerman}{\mbox{\cite{Gibbs2Reprint1+Gibbs2Englisch,NeumannEnglisch,Bridgman,LyuboshitzPodgoretskiiEJP,DenbighRedheadEJP}}.}
The GP2 is not the subject of the present paper.

In what follows, it will be crucial to understand the meanings of
the terms \textit{\mbox{\textit{(non-)identical}}} and \textit{(in)distinguishable}.
Roughly speaking, two particles are called identical if they agree
in all their permanent properties (such as electric charge), and they
are called indistinguishable if interchanging them does not alter
any microstate. Clearly, non-identical particles are always distinguishable
(even if they are very similar). Identical \textit{quantum} particles
are indistinguishable due to the symmetrization postulate \cite{MessiahGreenbergEJP}.
Identical \textit{classical} particles, by contrast, can be assumed
to be either distinguishable or indistinguishable; the two possibilities
turn out to be equivalent (cf. section~5.6 in \cite{Peters2010EJP}).
For a more thorough discussion see section~2 in \cite{Peters2010EJP}.

The GP1 is usually demonstrated and resolved as follows (e.g., in
\cite{SchroedingerEnglisch,HuangEnglisch,Pathria}). Two ideal gases
of the same kind with equal particle number, volume, and temperature
are combined. (Two gases, each consisting of identical particles,
are said to be \textit{of the same kind} if the particles of one
gas are identical to the particles of the other gas.) If the particles
of the two gases are pairwise distinguishable (which implies that
they are classical), then one calculates an entropy increase from
the combining process. This entropy increase is considered paradoxical.
If, on the contrary, the particles of the two gases are indistinguishable,
then one calculates a zero entropy change, as there ought to be.
On this account, quantum mechanics, in which identical particles
are necessarily indistinguishable, is claimed to be the resolution
of the GP1. 

Both the demonstration and the resolution just outlined are unsatisfactory.
The demonstration is unsatisfactory because it is not clear why,
for pairwise distinguishable particles, an entropy increase during
the course of combination is considered paradoxical. If, for example,
the first gas is composed of the particles 1 to $N$ and the second
gas of the particles $N+1$ to $2N$, then the calculated entropy
increase can be ascribed to the fact that, after combination, the
particles 1 to $N$ can access the volume of the second gas and, correspondingly,
the particles $N+1$ to $2N$ can access the volume of the first gas.
The resolution is unsatisfactory for two reasons. Firstly, one can
conceive gases consisting of very similar but still pairwise non-identical
particles. In statistical-mechanical calculations, such gases can
approximately be treated as  if their particles were identical but
still pairwise distinguishable. As a consequence, these gases suffer
from the GP1, even in quantum mechanics. (An example of such a gas,
along with a quantum-mechanically valid demonstration of the GP1,
is given in section~\ref{sec:Demonstration-des-GP1}.) Secondly,
the resolution above implies that the mere concept of distinguishable
identical particles is at odds with thermodynamics \cite{HestenesEJP}.
In the opinion of the present author such a connection between these
otherwise unrelated subjects would be rather surprising, to say the
least.

In section~\ref{sec:Demonstration-des-GP1}, the GP1 is recast in
a form which shows that, for one thing, the GP1 is indeed paradoxical
(by contradicting the second law of thermodynamics) and, for another
thing, the GP1 also affects \textit{quantum} statistical mechanics.
Instead of combining two gases, a single gas is partitioned.
Thus, instead of an entropy increase, one calculates an entropy decrease
which poses a clear contradiction to the second law \cite{Swendsen2006EJP}.
Furthermore, instead of considering a gas of pairwise distinguishable
identical (and thus, necessarily classical) particles, a gas of pairwise
non-identical but very similar particles is considered, allowing the
particles to be quantum-mechanical.

In section~\ref{sec:Physikalischer-Ursprung-und}, the origin of
the GP1 is identified and its resolution presented. It is shown
that, when calculating the entropy of a system of distinguishable
particles, one generally has to take account of two effects: 
\begin{enumerate}
\item If there is an uncertainty about the system's particle composition
(i.e., about which particles are located in the system), then this
uncertainty contributes to the system's entropy.
\item If the system consists of subsystems and there are particles which
possess a non-zero probability of presence in more than one subsystem,
then these subsystems are correlated (due to the fact that if such
a particle is located in one subsystem, it cannot be located in another).
If such correlations exist, then the system's entropy is smaller than
the sum of the entropies of the subsystems.
\end{enumerate}
These two effects resolve the GP1. As a matter of fact, they are
fundamental to the statistical-mechanical treatment of systems of
distinguishable particles, be it systems of non-identical particles
(cf. section~4 in \cite{Peters2010EJP}) or systems of distinguishable
identical classical particles (cf. section~5 in \cite{Peters2010EJP}).

In section~\ref{sec:ClassicalResolution}, the GP1 in its unaltered
original form is resolved by applying the insights of section~\ref{sec:Physikalischer-Ursprung-und}
to systems of distinguishable identical classical particles.

The present paper is accessible to undergraduates familiar with
the concepts of statistical mechanics. It is particularly aimed at
university teachers and textbook writers. \vspace{1ex}
\\
\textsc{Remark:}\smallskip{}
\\
$\star$ Some authors ``resolve'' the GP1 by using tailored entropy
expressions that do not agree with the standard definition used in
statistical mechanics \cite{Swendsen2002RedefiningEntropyEJP,JaynesRedefiningEntropyEJP}.
Others dismiss the correspondence between statistical-mechanical and
thermodynamical entropy (in order to allow violations of the statistical-mechanical
version of the second law) \cite{VersteeghDieksEJP}. The present
resolution, by contrast, adheres (for quantum systems) to the definition
\begin{equation}
S=-k\sum_{\textrm{all microstates }\mathsf{m}}P(\mathsf{m})\:\ln\, P(\mathsf{m})\label{eq:Entropiedefinition}
\end{equation}
{[}$k$ is Boltzmann's constant, $P(\mathsf{m})$ denotes the probability
of the microstate $\mathsf{m}${]} and does not question the correspondence
between statistical-mechanical and thermodynamical entropy.

\section{Demonstration of the GP1\label{sec:Demonstration-des-GP1}}

Consider a C$_{60}$ fullerene (buckyball) $\mathfrak{p}$ made up
of 30 carbon-12 atoms and 30 carbon-13 atoms. Let $\mathfrak{p}$
be located in a vessel of volume $V$ at temperature $T$. The (canonical)
partition function of this one-particle system has the form%
\begin{equation}
z(T,V,\mathfrak{p})=z_{\textrm{t.m.}}(T,V,\mathfrak{p})\, z_{\textrm{i.s.}}(T,\mathfrak{p})\label{eq:ZeT(oBsT)}
\end{equation}
\cite{rushbrookeFaktorisierteZustandssumme}, where $z_{\textrm{t.m.}}(T,V,\mathfrak{p})$
denotes the partition function for $\mathfrak{p}$'s translational
motion and $z_{\textrm{i.s.}}(T,\mathfrak{p})$ denotes the partition
function for $\mathfrak{p}$'s internal structure. As is well known,%
\begin{equation}
z_{\textrm{t.m.}}(T,V,\mathfrak{p})=V\left(\frac{mkT}{2\pi\hbar^{2}}\right)^{\frac{3}{2}}\label{eq:ZdTb}
\end{equation}
\cite{rushbrookeZustandssummeNurQMEJP} ($m$ denotes the mass of
$\mathfrak{p}$). $z_{\textrm{i.s.}}(T,\mathfrak{p})$ primarily
stems from $\mathfrak{p}$'s vibrational and rotational degrees of
freedom and is independent of $V$.

There are billions of non-equivalent ways to form a buckyball from
30 carbon-12 atoms and 30 carbon-13 atoms. All these buckyballs differ
from one another by the arrangement of the carbon isotopes and, as
a consequence, are pairwise non-identical. Now, let $2N$ $\left(N\gg1\right)$
of these buckyballs, $\mathfrak{p}_{1},\,\mathfrak{p}_{2},\,\ldots,\,\mathfrak{p}_{2N}$,
be confined to a vessel of volume $2V$ at temperature $T$. For a
sufficiently large volume, $\mathfrak{p}_{1},\,\mathfrak{p}_{2},\,\ldots,\,\mathfrak{p}_{2N}$
form a (nearly) ideal gas $\mathbb{S}_{0}$ with the partition function%
\begin{eqnarray}
Z_{0}(T,2V,\mathfrak{p}_{1},\mathfrak{p}_{2},\ldots,\mathfrak{p}_{2N})\; & \overset{\textrm{ideal gas}}{=} & \;\underset{n=1}{\overset{2N}{\prod}}z(T,2V,\mathfrak{p}_{n})\nonumber \\
 & \overset{\eqref{eq:ZeT(oBsT)}}{\underset{\hphantom{\hphantom{\textrm{ideal gas}}}}{=}} & \;\underset{n=1}{\overset{2N}{\prod}}z_{\textrm{t.m.}}(T,2V,\mathfrak{p}_{n})\, z_{\textrm{i.s.}}(T,\mathfrak{p}_{n})\:.\label{eq:HZiG(bbT)}
\end{eqnarray}
Since $\mathfrak{p}_{1},\,\mathfrak{p}_{2},\,\ldots,\,\mathfrak{p}_{2N}$
all have the same mass $m$, the partition functions for their translational
motions agree: 
\begin{equation}
z_{\textrm{t.m.}}(T,2V,\mathfrak{p}_{1})=\ldots=z_{\textrm{t.m.}}(T,2V,\mathfrak{p}_{2N})\overset{\eqref{eq:ZdTb}}{=}2V\left(\frac{mkT}{2\pi\hbar^{2}}\right)^{\frac{3}{2}}\:.\label{eq:Gleiches_z_t.m.}
\end{equation}
The partition functions for their internal structures, by contrast,
only agree \textit{approximately} because a buckyball's arrangement
of carbon isotopes slightly affects its vibrational modes and moments
of inertia. In order not to base the present demonstration of the
GP1 on approximations relying on the similarity between $\mathfrak{p}_{1},\,\mathfrak{p}_{2},\,\ldots,\,\mathfrak{p}_{2N}$,
let the temperature $T$ be so low that the internal degrees of freedom
of $\mathfrak{p}_{1},\,\mathfrak{p}_{2},\,\ldots,\,\mathfrak{p}_{2N}$
are frozen out. (The volume $2V$ is assumed so large that the buckyballs
still can be treated as an ideal gas.) Furthermore, let $\mathfrak{p}_{1},\,\mathfrak{p}_{2},\,\ldots,\,\mathfrak{p}_{2N}$
be chosen such that they all have the same ground-state degeneracy
$g$. Under these conditions, the partition functions for their
internal structures become
\begin{equation}
z_{\textrm{i.s.}}(T,\mathfrak{p}_{n})=g\exp\left(-\frac{\epsilon_{\mathfrak{p}_{n}}}{kT}\right)\qquad n=1,\,2,\,\ldots,\,2N\:,\label{eq:zisfrozen}
\end{equation}
where $\epsilon_{\mathfrak{p}_{n}}$ denotes the ground-state energy
of $\mathfrak{p}_{n}$. Finally, setting each buckyball's ground-state
energy to $0$ eliminates the remaining differences between these
partition functions and yields

\begin{equation}
z_{\textrm{i.s.}}(T,\mathfrak{p}_{1})=\ldots=z_{\textrm{i.s.}}(T,\mathfrak{p}_{2N})\overset{\eqref{eq:zisfrozen}}{=}g\:.\label{eq:Aehnliches_z_i.s.}
\end{equation}
With this, $\mathbb{S}_{0}$'s partition function becomes%
\begin{equation}
Z_{0}(T,2V,2N)\;\overset{\eqref{eq:HZiG(bbT)},\eqref{eq:Gleiches_z_t.m.},\eqref{eq:Aehnliches_z_i.s.}}{=}\;\left[2V\left(\frac{mkT}{2\pi\hbar^{2}}\right)^{\frac{3}{2}}g\right]^{2N}\:.\label{eq:ZiG(oBsT)}
\end{equation}

As a side note, \eqref{eq:ZiG(oBsT)} has the form of a partition
function of an ideal gas of $2N$ \textit{distinguishable identical}
particles: For one thing, there is no indistinguishability factor
$1/\!\left(2N\right)!$ in \eqref{eq:ZiG(oBsT)}; for another thing,
\eqref{eq:ZiG(oBsT)} does not depend on the particle identities
$\mathfrak{p}_{1},\,\mathfrak{p}_{2},\,\ldots,\,\mathfrak{p}_{2N}$,
but only on the particle number $2N$. Thus, as noted in section \ref{sec:Einleitung},
one may expect that the GP1 arises.

The entropy $\left.S(T,\ldots)\right.$ of a system in canonical equilibrium
results from its partition function $\left.Z(T,\ldots)\right.$ by%
\begin{equation}
S(T,\ldots)=\frac{\partial}{\partial T}\left[kT\ln Z(T,\ldots)\right]\label{eq:EaZ}
\end{equation}
\cite{Feynman}. Applying \eqref{eq:EaZ} to \eqref{eq:ZiG(oBsT)}
yields $\mathbb{S}_{0}$'s entropy%
\begin{equation}
S_{0}(T,2V,2N)=2Nk\left\{ \ln\left(2V\right)+\ln\left[\left(\frac{mkT}{2\pi\hbar^{2}}\right)^{\frac{3}{2}}g\right]+\frac{3}{2}\right\} \:.\label{eq:EiG(oBsT)}
\end{equation}

Inserting a partition in the middle of $\mathbb{S}_{0}$ divides $\mathbb{S}_{0}$
into two equal subsystems $\mathbb{S}_{1}$ and $\mathbb{S}_{2}$,
each with volume $V$ and each containing $N$ buckyballs. Since $\mathbb{S}_{1}$
and $\mathbb{S}_{2}$ are ideal gases consisting of the same particles
as $\mathbb{S}_{0}$, their partition functions are%
\begin{equation}
Z_{1}(T,{\textstyle V},{\textstyle N})=Z_{2}(T,{\textstyle V},{\textstyle N})=Z_{0}(T,{\textstyle V},{\textstyle N})\label{eq:ZustdTEils(fehlerhaft)}
\end{equation}
yielding the entropies%
\begin{equation}
S_{1}(T,{\textstyle V},N)=S_{2}(T,{\textstyle V},N)=S_{0}(T,{\textstyle V},N).\label{eq:EntrdTeils(fehlerhaft)}
\end{equation}
Thus, the entropy of the partitioned total system $\mathbb{S}_{1+2}$
consisting of $\mathbb{S}_{1}$ and $\mathbb{S}_{2}$ is%
\begin{equation}
S_{1+2}(T,2V,2N)=S_{1}(T,{\textstyle V},N)+S_{2}(T,{\textstyle V},{\textstyle N})\overset{\eqref{eq:EntrdTeils(fehlerhaft)}}{=}2S_{0}(T,{\textstyle V},N).\label{eq:EuS1+2}
\end{equation}
Comparison with the entropy of the original (unpartitioned) system
$\mathbb{S}_{0}$ shows that the partitioning decreases the entropy
by%
\begin{eqnarray}
S_{0}(T,2V,2N)-S_{1+2}(T,2V,2N) & \overset{\eqref{eq:EuS1+2}}{=} & S_{0}(T,2V,2N)-2S_{0}(T,{\textstyle V},{\textstyle N})\nonumber \\
 & \overset{\eqref{eq:EiG(oBsT)}}{\underset{\hphantom{\eqref{eq:EuS1+2}}}{=}} & 2Nk\ln2\:.\label{eq:EadU}
\end{eqnarray}
Since the partitioning does not perform work (see remark below),
and since $\mathbb{S}_{0}$'s energy equals $\mathbb{S}_{1+2}$'s
energy (see appendix \hyperref[AppendixA]{A}), there is, according
to the first law of thermodynamics, no heat transfer to the thermal
reservoir. Hence, the entropy decrease \eqref{eq:EadU} contradicts
the second law of thermodynamics.\vspace{1ex}
\\
\textsc{Remark:}\smallskip{}
\\
$\star$ The claim that the partitioning does not perform work is
taken for granted here. Strictly speaking, inserting a partition in
$\mathbb{S}_{0}$ involves a tiny amount of work (because the buckyballs'
wave functions are deformed) and it must be shown that this amount
is negligible compared to $2NkT\ln2$.

\section{Origin and resolution of the GP1\label{sec:Physikalischer-Ursprung-und}}

Two flaws are responsible for the false entropy decrease \eqref{eq:EadU}
calculated in the previous section. One is located in \eqref{eq:ZustdTEils(fehlerhaft)},
the other in \eqref{eq:EuS1+2}. Before pointing out and correcting
these flaws, it is worthwhile to take a closer look at the term \textit{uncertainty}
and its use in the context of the present situation.

Each microstate of a system in canonical equilibrium possesses a certain
non-zero probability. Hence, if the microstates of such a system differ
in a specific property, then there is an \textit{uncertainty} about
this property. For example, in $\mathbb{S}_{0}$ there is, classically
speaking, an uncertainty about the position of particle $\mathfrak{p}_{1}$
because there are microstates in which $\mathfrak{p}_{1}$ is located
in the first half of the vessel and other microstates in which $\mathfrak{p}_{1}$
is located in the second half. Accordingly, after partitioning, there
is an uncertainty in $\mathbb{S}_{1+2}$ about whether $\mathfrak{p}_{1}$
is located in $\mathbb{S}_{1}$ or $\mathbb{S}_{2}$. From $\mathbb{S}_{1}$'s
perspective, there is an uncertainty about whether $\mathfrak{p}_{1}$
is located within the system \textit{at all}. Since the same uncertainty
exists with regard to the other particles $\mathfrak{p}_{2},\,\mathfrak{p}_{3},\,\ldots,\,\mathfrak{p}_{2N}$,
there is an uncertainty about the \textit{particle composition} of
$\mathbb{S}_{1}$, i.e., the set of particles located in $\mathbb{S}_{1}$.
In short, the uncertainty about the particle positions in $\mathbb{S}_{0}$
results, after partitioning, in an uncertainty about which particles
are located in $\mathbb{S}_{1}$. 

By definition, there is no uncertainty about $\mathbb{S}_{0}$'s particle
composition $\{\mathfrak{p}_{1},\,\mathfrak{p}_{2},\,\ldots,\,\mathfrak{p}_{2N}\}$.
In contrast, as just explained, $\mathbb{S}_{1}$'s particle composition
might be $\{\mathfrak{p}_{1},\,\mathfrak{p}_{2},\,\ldots,\,\mathfrak{p}_{N}\}$
in one microstate and, e.g., $\{\mathfrak{p}_{2},\,\mathfrak{p}_{3},\,\ldots,\,\mathfrak{p}_{N+1}\}$
in another. Provided that $\mathbb{S}_{1}$ consists of $N$ particles
(cf.~second remark below), there are ${2N \choose N}$ possible particle
compositions $\zeta_{1},\,\zeta_{2},\,\ldots,\,\zeta_{{2N \choose N}}$
of $\mathbb{S}_{1}$, namely, all $N$-element subsets of $\mathbb{S}_{0}$'s
particle composition: 
\begin{equation}
\zeta_{1},\,\zeta_{2},\,\ldots,\,\zeta_{{2N \choose N}}\subset\left\{ \mathfrak{p}_{1},\,\mathfrak{p}_{2},\,\ldots,\,\mathfrak{p}_{2N}\right\} \label{eq:Zeta-Teilmengen}
\end{equation}
\begin{equation}
\left|\zeta_{1}\right|=\left|\zeta_{2}\right|=\ldots=|\zeta_{{2N \choose N}}|=N\;.\label{eq:Zeta-Kardinalitaet}
\end{equation}

Now, consider the hypothetical system $\mathbb{S}_{1}^{\zeta}$ which
is thought to be in all respects the same as $\mathbb{S}_{1}$ except
that there is no uncertainty about its particle composition~$\left.\zeta\in\{\zeta_{1},\,\zeta_{2},\,\ldots,\,\zeta_{{2N \choose N}}\}\right.$.
The partition function of $\mathbb{S}_{1}^{\zeta}$ is
\begin{eqnarray}
Z_{1}\left(T,{\textstyle V},{\textstyle N}\mid\zeta\right)\: & \overset{\textrm{ideal gas}}{\underset{\hphantom{\eqref{eq:Zeta-Teilmengen},\eqref{eq:Gleiches_z_t.m.},\eqref{eq:Aehnliches_z_i.s.}}}{=}} & \:\:\underset{\mathfrak{q}\in\zeta}{\prod}z(T,{\textstyle V},\mathfrak{q})\nonumber \\
 & \overset{\eqref{eq:ZeT(oBsT)}}{\underset{\hphantom{\eqref{eq:Zeta-Teilmengen},\eqref{eq:Gleiches_z_t.m.},\eqref{eq:Aehnliches_z_i.s.}}}{=}} & \:\:\underset{\mathfrak{q}\in\zeta}{\prod}z_{\textrm{t.m.}}(T,{\textstyle V},\mathfrak{q})\, z_{\textrm{i.s.}}(T,\mathfrak{q})\nonumber \\
 & \overset{\eqref{eq:Zeta-Teilmengen},\eqref{eq:Gleiches_z_t.m.},\eqref{eq:Aehnliches_z_i.s.}}{=} & \:\:\underset{\mathfrak{q}\in\zeta}{\prod}V\left(\frac{mkT}{2\pi\hbar^{2}}\right)^{\frac{3}{2}}g\nonumber \\
 & \overset{\eqref{eq:Zeta-Kardinalitaet},\eqref{eq:ZiG(oBsT)}}{\underset{\hphantom{\eqref{eq:Zeta-Teilmengen},\eqref{eq:Gleiches_z_t.m.},\eqref{eq:Aehnliches_z_i.s.}}}{=}} & \:\: Z_{0}(T,{\textstyle V},{\textstyle N})\:.\label{eq:Z1(T,V,Zeta_i)}
\end{eqnarray}
This shows that setting $\mathbb{S}_{1}$'s partition function equal
to $\mathbb{S}_{0}$'s partition function, as was done in \eqref{eq:ZustdTEils(fehlerhaft)},
amounts to ignoring the uncertainty about $\mathbb{S}_{1}$'s particle
composition. The correct calculation of $\mathbb{S}_{1}$'s partition
function, by contrast, takes all possible particle compositions of
$\mathbb{S}_{1}$ into account:%
\begin{eqnarray}
Z_{1}(T,{\textstyle V},N) & \underset{\hphantom{\eqref{eq:Z1(T,V,Zeta_i)}}}{=}\, & \underset{i=1}{\overset{{2N \choose N}}{\sum}}\;\;\underset{\textrm{of }\mathbb{S}_{1}\textrm{ is }\zeta_{i}}{\underset{\textrm{particle composition}}{\underset{\textrm{ of }\mathbb{S}_{1}\textrm{ in which the}}{\underset{\textrm{all microstates }\mathsf{m}}{\sum}}}}\;\exp\left(-\frac{E_{\mathsf{m}}}{kT}\right)\nonumber \\
 & \underset{\hphantom{\eqref{eq:Z1(T,V,Zeta_i)}}}{=}\, & \underset{i=1}{\overset{{2N \choose N}}{\sum}}\;\; Z_{1}\left(T,{\textstyle V},N\mid\zeta_{i}\right)\nonumber \\
 & \overset{\eqref{eq:Z1(T,V,Zeta_i)}}{=}\, & {2N \choose N}Z_{0}(T,{\textstyle V},N)\label{eq:ZustdTeils}
\end{eqnarray}
($E_{\mathsf{m}}$ denotes the energy of the microstate $\mathsf{m}$).
From \eqref{eq:ZustdTeils}, one obtains $\mathbb{S}_{1}$'s entropy%
\begin{equation}
S_{1}(T,{\textstyle V},N)\overset{\eqref{eq:EaZ}}{=}S_{0}(T,{\textstyle V},N)+k\ln{2N \choose N}\:.\label{eq:EntrvS1}
\end{equation}
Compared to \eqref{eq:EntrdTeils(fehlerhaft)}, there is an additional
term $k\ln{\textstyle {2N \choose N}}$ reflecting the uncertainty
about the particle composition of $\mathbb{S}_{1}$.

Applying the above reasoning to $\mathbb{S}_{2}$ yields%
{} 
\begin{equation}
S_{2}(T,{\textstyle V},{\textstyle N})=S_{1}(T,{\textstyle V},N)\:,\label{eq:EntrvS2}
\end{equation}
as required by symmetry. 

Each microstate of the partitioned total system $\mathbb{S}_{1+2}$
is the tensor product $\mathsf{m}\otimes\mathsf{n}$ of a microstate
$\mathsf{m}$ of $\mathbb{S}_{1}$ and a microstate $\mathsf{n}$
of $\mathbb{S}_{2}$. Since there is no interaction between $\mathbb{S}_{1}$
and $\mathbb{S}_{2}$,
\begin{equation}
E_{\mathsf{\mathsf{m}\otimes\mathsf{n}}}=E_{\mathsf{\mathsf{m}}}+E_{\mathsf{\mathsf{n}}}\:.\label{eq:AdditiveEnergie}
\end{equation}
However, not every microstate $\mathsf{n}$ of $\mathbb{S}_{2}$ is
compatible with every microstate $\mathsf{m}$ of $\mathbb{S}_{1}$.
If in $\mathsf{m}$ a certain particle is located in $\mathbb{S}_{1}$,
and in $\mathsf{n}$ the same particle is located in $\mathbb{S}_{2}$,
then the tensor product $\mathsf{m}\otimes\mathsf{n}$ is not well-defined
because the same particle appears twice in it. On the other hand,
if in $\mathsf{m}$ a certain particle out of $\mathfrak{p}_{1},\,\mathfrak{p}_{2},\,\ldots,\,\mathfrak{p}_{2N}$
is \textit{not} located in $\mathbb{S}_{1}$, and in $\mathsf{n}$
the same particle is not located in $\mathbb{S}_{2}$, then $\mathsf{m}\otimes\mathsf{n}$
is inadmissible for $\mathbb{S}_{1+2}$ because it does not include
all particles of $\mathbb{S}_{1+2}$. Thus, for $\mathsf{n}$ to be
compatible with $\mathsf{m}$, the particle composition of $\mathbb{S}_{2}$
in $\mathsf{n}$ must be complementary to the particle composition
of $\mathbb{S}_{1}$ in $\mathsf{m}$. For example, if the particle
composition of $\mathbb{S}_{1}$ in $\mathsf{m}$ is $\{\mathfrak{p}_{1},\,\mathfrak{p}_{2},\,\ldots,\,\mathfrak{p}_{N}\}$,
then the particle composition of $\mathbb{S}_{2}$ in a microstate
$\mathsf{n}$ compatible with $\mathsf{m}$ must be $\{\mathfrak{p}_{N+1},\,\mathfrak{p}_{N+2},\,\ldots,\,\mathfrak{p}_{2N}\}$.
If this complementary relationship between compatible microstates
of $\mathbb{S}_{1}$ and $\mathbb{S}_{2}$ is taken into account,
one obtains
\begin{eqnarray}
\fl Z_{1+2}(T,2V,2N) & \overset{\hphantom{\eqref{eq:AdditiveEnergie}}}{=}\, & \underset{i=1}{\overset{{2N \choose N}}{\sum}}\;\underset{\textrm{of }\mathbb{S}_{1}\textrm{ is }\zeta_{i}}{\underset{\textrm{particle composition}}{\underset{\textrm{ of }\mathbb{S}_{1}\textrm{ in which the}}{\underset{\textrm{all microstates }\mathsf{m}}{\sum}}}}\;\;\;\;\underset{}{\underset{\textrm{compatible with }\mathsf{m}}{\underset{\textrm{ of }\mathbb{S}_{2}\textrm{ which are }}{\underset{\textrm{all microstates }\mathsf{n}}{\sum}}}}\;\exp\left(-\frac{E_{\mathsf{\mathsf{m}\otimes\mathsf{n}}}}{kT}\right)\nonumber \\
\fl\vphantom{Z_{1+2}(T,2V,2N)} & \overset{\eqref{eq:AdditiveEnergie}}{\underset{}{=}}\, & \underset{i=1}{\overset{{2N \choose N}}{\sum}}\;\underset{\textrm{of }\mathbb{S}_{1}\textrm{ is }\zeta_{i}}{\underset{\textrm{particle composition}}{\underset{\textrm{ of }\mathbb{S}_{1}\textrm{ in which the}}{\underset{\textrm{all microstates }\mathsf{m}}{\sum}}}}\exp\left(-\frac{E_{\mathsf{\mathsf{m}}}}{kT}\right)\underset{\textrm{of }\mathbb{S}_{2}\textrm{ is }\zeta_{i}^{\mathsf{c}}}{\underset{\textrm{particle composition}}{\underset{\textrm{ of }\mathbb{S}_{2}\textrm{ in which the}}{\underset{\textrm{all microstates }\mathsf{n}}{\sum}}}}\exp\left(-\frac{E_{\mathsf{\mathsf{n}}}}{kT}\right)\nonumber \\
\fl\vphantom{Z_{1+2}(T,2V,2N)} & \overset{\hphantom{\eqref{eq:AdditiveEnergie}}}{=}\, & \underset{i=1}{\overset{{2N \choose N}}{\sum}}\; Z_{1}\left(T,{\textstyle V},{\textstyle N}\mid\zeta_{i}\right)\; Z_{2}\left(T,{\textstyle V},N\mid\zeta_{i}^{\mathsf{c}}\right)\:,\label{eq:ZduGesVorl}
\end{eqnarray}
where $\zeta_{i}^{\mathsf{c}}$ denotes the particle composition that
is complementary to $\zeta_{i}$ and $Z_{2}\left(T,{\textstyle V},{\textstyle N}\mid\zeta_{i}^{\mathsf{c}}\right)$
denotes the partition function of a (hypothetical) system which is
in all respects the same as $\mathbb{S}_{2}$ except that its particle
composition is $\zeta_{i}^{\mathsf{c}}\hspace{0bp}$. As in $\eqref{eq:Z1(T,V,Zeta_i)}$,
one has 
\begin{equation}
Z_{2}\left(T,{\textstyle V},{\textstyle N}\mid\zeta_{i}^{\mathsf{c}}\right)=Z_{0}(T,{\textstyle V},N)\:.\label{eq:Z2(T,V,Zeta_i_c)}
\end{equation}
With this, $\mathbb{S}_{1+2}$'s partition function becomes
\begin{equation}
Z_{1+2}(T,2V,2N)\overset{\eqref{eq:ZduGesVorl},\eqref{eq:Z1(T,V,Zeta_i)},\eqref{eq:Z2(T,V,Zeta_i_c)}}{=}{2N \choose N}\;\left[Z_{0}(T,{\textstyle V},{\textstyle N})\right]^{2}\label{eq:ZduGes}
\end{equation}
yielding the entropy%
\begin{equation}
S_{1+2}(T,2V,2N)\overset{\eqref{eq:EaZ},\eqref{eq:ZduGes}}{=}2S_{0}(T,{\textstyle V},N)+k\ln{2N \choose N}\:.\label{eq:EntrduGes}
\end{equation}
$S_{1+2}(T,2V,2N)$ is smaller than
\begin{equation}
S_{1}(T,{\textstyle V},{\textstyle N})+S_{2}(T,{\textstyle V},N)\overset{\eqref{eq:EntrvS1},\eqref{eq:EntrvS2}}{=}2\left[S_{0}(T,{\textstyle V},N)+k\ln{2N \choose N}\right]\label{eq:S1(T,V,N)}
\end{equation}
by $k\ln{\textstyle {2N \choose N}}$, reflecting the fact that $\mathbb{S}_{1}$
and $\mathbb{S}_{2}$ are correlated by their complementary particle
compositions. This correlation was ignored in section \ref{sec:Demonstration-des-GP1}
when, in \eqref{eq:EuS1+2}, the entropy of $\mathbb{S}_{1+2}$ was
set to the sum of the entropies of $\mathbb{S}_{1}$ and $\mathbb{S}_{2}$.

Comparison of the entropies of $\mathbb{S}_{0}$ and $\mathbb{S}_{1+2}$
shows that the partitioning does not change the entropy:%
\begin{eqnarray}
\fl S_{0}(T,2V,2N)-S_{1+2}(T,2V,2N)\; & \overset{\eqref{eq:EntrduGes}}{\underset{\hphantom{\textrm{Stirling}}}{=}} & \;\; S_{0}(T,2V,2N)-2S_{0}(T,{\textstyle V},N)-k\ln{2N \choose N}\nonumber \\
 & \overset{\eqref{eq:EiG(oBsT)}}{\underset{\hphantom{\textrm{Stirling}}}{=}} & \;\;2Nk\ln2-k\left[\ln\left(2N\right)!-2\ln{\textstyle N}!\right]\nonumber \\
 & \overset{\textrm{Stirling}}{\approx} & \;\;0\:.\label{eq:KEAe}
\end{eqnarray}
Thus, the GP1 is resolved.\vspace{1ex}
\\
\textsc{Remarks:}\vspace*{\smallskipamount}
\\
$\star$ Traditionally, the GP1 is demonstrated by combining two gas
systems and not by partitioning a single one. In order to resolve
the GP1 in its traditional form, it is crucial to carefully specify
the initial bipartite system. If this initial system is, e.g., $\mathbb{S}_{1+2}$,
then combining its subsystems $\mathbb{S}_{1}$ and $\mathbb{S}_{2}$
yields $\mathbb{S}_{0}$ and, according to \eqref{eq:KEAe}, a zero
entropy change. If, on the other hand, the initial system consists
of, e.g., two subsystems $\mathbb{S}_{3}$ and $\mathbb{S}_{4}$,
where $\mathbb{S}_{3}$'s particle composition is $\{\mathfrak{p}_{1},\,\mathfrak{p}_{2},\,\ldots,\,\mathfrak{p}_{N}\}$
and $\mathbb{S}_{4}$'s particle composition is $\{\mathfrak{p}_{N+1},\,\mathfrak{p}_{N+2},\,\ldots,\,\mathfrak{p}_{2N}\}$,
then combining $\mathbb{S}_{3}$ and $\mathbb{S}_{4}$ increases the
entropy. However, as with two gases of different kinds, this entropy
increase is not paradoxical; it reflects the fact that the combining
process cannot be reversed by reinserting the partition, since, after
reinsertion, there is no longer certainty that the particle compositions
of the subsystems are $\{\mathfrak{p}_{1},\,\mathfrak{p}_{2},\,\ldots,\,\mathfrak{p}_{N}\}$
and $\{\mathfrak{p}_{N+1},\,\mathfrak{p}_{N+2},\,\ldots,\,\mathfrak{p}_{2N}\}$,
respectively.\smallskip{}

\noindent $\star$ The assumed certainty about $\mathbb{S}_{0}$'s
particle composition $\{\mathfrak{p}_{1},\,\mathfrak{p}_{2},\,\ldots,\,\mathfrak{p}_{2N}\}$
is a special case. In the general case, where there is an uncertainty
about $\mathbb{S}_{0}$'s and thus also about $\mathbb{S}_{1+2}$'s
particle composition, the particle compositions of $\mathbb{S}_{1}$
and $\mathbb{S}_{2}$ need no longer be complementary but only disjoint
(cf. section 4.3.2 in \cite{Peters2010EJP}).\smallskip{}

\noindent $\star$ In addition to the uncertainty about \emph{which}
particles are located in $\mathbb{S}_{1}$ and  $\mathbb{S}_{2}$,
respectively, there is also an uncertainty about \emph{how many} particles
are located in each subsystem. However, the latter uncertainty contributes
only negligibly to the entropies of $\mathbb{S}_{1}$, $\mathbb{S}_{2}$,
and $\mathbb{S}_{1+2}$ \cite{Tolman,CasperFreierEJP}. It is therefore
acceptable to disregard this uncertainty by assuming the particle
number $N$ for $\mathbb{S}_{1}$ and $\mathbb{S}_{2}$. As an aside,
this disregard is responsible for the small discrepancy between $S_{0}(T,2V,2N)$
and $S_{1+2}(T,2V,2N)$ expressed by the Stirling approximation in
\eqref{eq:KEAe}. \smallskip{}

\noindent $\star$ The concept of the canonical partition function
is based on the assumption that the probability of any microstate
$\mathsf{m}$ of a system in canonical equilibrium is given by
\begin{equation}
P(\mathsf{m})=C\exp\left(-\frac{E_{\mathsf{m}}}{kT}\right)\:,\label{eq:P(m)/P(n)}
\end{equation}
where the normalization constant $C$ may depend on the equilibrium
macrostate, but not on the microstate $\mathsf{m}$. Non-ergodic systems,
such as $\mathbb{S}_{1}$, generally do not satisfy this assumption.
($\mathbb{S}_{1}$ is non-ergodic because a microstate $\mathsf{m}$
cannot dynamically evolve to a microstate $\mathsf{n}$ if $\mathbb{S}_{1}$'s
particle composition in $\mathsf{m}$ differs from that in $\mathsf{n}$.)
Therefore, using $\mathbb{S}_{1}$'s partition function \eqref{eq:ZustdTeils}
requires justification. This justification is provided in appendix
\hyperref[AppendixB]{B}. (The justifications for using the partition
functions of $\mathbb{S}_{2}$ and $\mathbb{S}_{1+2}$ are similar.)

\section{Resolution of the GP1 for distinguishable identical classical particles\label{sec:ClassicalResolution}}

Consider a structureless classical particle $\mathfrak{q}_{1}$ of
mass $m$ confined to a vessel of volume $V$ at temperature $T$.
Assuming an elementary phase space volume of $\left(2\pi\hbar\right)^{3}$,
the partition function of this one-particle system is
\begin{equation}
z(T,V,\mathfrak{q}_{1})=V\left(\frac{mkT}{2\pi\hbar^{2}}\right)^{\frac{3}{2}}\label{eq:klZISingleNon-phsical}
\end{equation}
\cite{rushbrookeZustandssummeNurClassicEJP}. Now, let the vessel
be filled with $N$ $\left(N\gg1\right)$ distinguishable identical
particles $\mathfrak{q}_{1},\,\mathfrak{q}_{2},\,\ldots,\,\mathfrak{q}_{N}$.
For a sufficiently large volume, $\mathfrak{q}_{1},\,\mathfrak{q}_{2},\,\ldots,\,\mathfrak{q}_{N}$
form a (nearly) ideal gas $\mathbb{S}_{\textrm{I}}$ with the partition
function
\begin{equation}
Z_{\textrm{I}}(T,V,\mathfrak{q}_{1},\,\mathfrak{q}_{2},\,\ldots,\,\mathfrak{q}_{N})=\left[z(T,V,\mathfrak{q}_{1})\right]^{N}\overset{\eqref{eq:klZISingleNon-phsical}}{=}\left[V\left(\frac{mkT}{2\pi\hbar^{2}}\right)^{\frac{3}{2}}\right]^{N}\:.\label{eq:klZIGasNon-phsical}
\end{equation}

Let $\mathfrak{q}_{1},\,\mathfrak{q}_{2},\,\ldots,\,\mathfrak{q}_{\mathfrak{N}}$
be all the particles in the universe that are identical to the particles
in $\mathbb{S}_{\textrm{I}}$ and assume $\mathfrak{N}\gg N$. Since,
by definition, $\mathfrak{q}_{1},\,\mathfrak{q}_{2},\,\ldots,\,\mathfrak{q}_{\mathfrak{N}}$
agree in all their permanent properties, there is, as a matter of
principle, no way to be sure that really $\mathfrak{q}_{1},\,\mathfrak{q}_{2},\,\ldots,\,\mathfrak{q}_{N}$
are located in $\mathbb{S}_{\textrm{I}}$ and not, e.g., $\mathfrak{q}_{2},\,\mathfrak{q}_{3},\,\ldots,\,\mathfrak{q}_{N+1}$.
Therefore, $\left\{ \mathfrak{q}_{1},\,\mathfrak{q}_{2},\,\ldots,\,\mathfrak{q}_{N}\right\} $
is just one of ${\mathfrak{N} \choose N}$ possible particle compositions
of $\mathbb{S}_{\textrm{I}}$. Taking this uncertainty about $\mathbb{S}_{\textrm{I}}$'s
particle composition into account results, as with \eqref{eq:ZustdTeils},
in an ${\mathfrak{N} \choose N}$ times larger partition function.
Thus, $\mathbb{S}_{\textrm{I}}$'s true partition function is
\begin{equation}
Z_{\textrm{I}}(T,V,N)\overset{\eqref{eq:klZIGasNon-phsical}}{=}{\mathfrak{N} \choose N}\left[V\left(\frac{mkT}{2\pi\hbar^{2}}\right)^{\frac{3}{2}}\right]^{N}\label{eq:klZI}
\end{equation}
yielding the entropy
\begin{equation}
S_{\textrm{I}}(T,{\textstyle V},N)\overset{\eqref{eq:EaZ},\eqref{eq:klZI}}{=}Nk\left(\ln V+\frac{3}{2}\ln\frac{mkT}{2\pi\hbar^{2}}+\frac{3}{2}\right)+k\ln{\mathfrak{N} \choose N}\:.\label{eq:klSI}
\end{equation}

Now, consider a second system $\mathbb{S}_{\textrm{II}}$ of volume
$V$ which is separated from $\mathbb{S}_{\textrm{I}}$ by a removable
partition. Let $\mathbb{S}_{\textrm{II}}$, at temperature $T$, contain
$N$ particles identical to the particles in $\mathbb{S}_{\textrm{I}}$.
Then, of course, the entropies of $\mathbb{S}_{\textrm{I}}$ and
$\mathbb{S}_{\textrm{II}}$ agree:
\begin{equation}
S_{\textrm{II}}(T,{\textstyle V},N)=S_{\textrm{I}}(T,{\textstyle V},N)\:.\label{eq:SIIgleichSI}
\end{equation}
Since $\mathfrak{N}\gg N$, $\mathbb{S}_{\textrm{I}}$ and $\mathbb{S}_{\textrm{II}}$
are nearly uncorrelated (see first remark below) and, as a consequence,
the entropy of the total system $\mathbb{S}_{\textrm{I+II}}$ is approximately
the sum of the entropies of its subsystems $\mathbb{S}_{\textrm{I}}$
and $\mathbb{S}_{\textrm{II}}$:
\begin{equation}
S_{\textrm{I+II}}(T,{\textstyle 2V},2N)\approx S_{\textrm{I}}(T,{\textstyle V},N)+S_{\textrm{II}}(T,{\textstyle V},N)\overset{\eqref{eq:SIIgleichSI}}{=}2S_{\textrm{I}}(T,{\textstyle V},N).\label{eq:SI+II}
\end{equation}

By removing the partition between $\mathbb{S}_{\textrm{I}}$ and $\mathbb{S}_{\textrm{II}}$,
one obtains the combined system $\mathbb{S}_{\textrm{I}\textrm{II}}$
with the entropy
\begin{equation}
S_{\textrm{III}}(T,{\textstyle 2V},2N)=S_{\textrm{I}}(T,2{\textstyle V},2N)\:.\label{eq:SIII}
\end{equation}
Comparison of the entropies of $\mathbb{S}_{\textrm{I+II}}$ and $\mathbb{S}_{\textrm{III}}$
shows that the combination does not change the entropy:
\begin{eqnarray}
\lefteqn{S_{\textrm{I}+\textrm{II}}(T,{\textstyle 2V},2N)-S_{\textrm{III}}(T,{\textstyle 2V},2N)\vphantom{{\mathfrak{N} \choose 2N}}}\;\nonumber \\
 &  & \quad\qquad\overset{\eqref{eq:SI+II},\eqref{eq:SIII}}{\underset{\hphantom{\textrm{Stirling}}}{\approx}}\;\;2S_{\textrm{I}}(T,{\textstyle V},N)-S_{\textrm{I}}(T,2{\textstyle V},2N)\vphantom{{\mathfrak{N} \choose 2N}}\nonumber \\
 &  & \quad\qquad\overset{\eqref{eq:klSI}}{\underset{\hphantom{\textrm{Stirling}}}{=}}\;\;2k\ln{\mathfrak{N} \choose N}-2Nk\ln2-k\ln{\mathfrak{N} \choose 2N}\nonumber \\
 &  & \quad\qquad\overset{\mathfrak{N}\gg N}{\underset{\hphantom{\textrm{Stirling}}}{\approx}}\;\; k\left[2\ln\mathfrak{N}^{N}-2\ln N!-2N\ln2-\ln\mathfrak{N}^{2N}+\ln\left(2N\right)!\right]\nonumber \\
 &  & \quad\qquad\overset{\textrm{Stirling}}{\approx}\;\;0\:.\label{eq:SI+IIgleichSIII}
\end{eqnarray}
\vspace{1ex}
\\
\textsc{Remarks:}\vspace*{\smallskipamount}
\\
$\star$ The resolution above is based on the assumption $\mathfrak{N}\gg N$.
The absence of the GP1 in this limiting case was found before by van
Kampen \cite{VanKampenEJP}. If the assumption $\mathfrak{N}\gg N$
is not satisfied, then the correlation between $\mathbb{S}_{\textrm{I}}$
and $\mathbb{S}_{\textrm{II}}$ arising from their disjoint particle
compositions is no longer negligible (see section~5.4 in \cite{Peters2010EJP}).
The general resolution which does not rely on $\mathfrak{N}\gg N$
is given in section~6.1.2 in \cite{Peters2010EJP}.\smallskip{}

\noindent $\star$ One might expect that $\mathfrak{N}$ is necessarily
an unknown and very large number and, as a consequence, that entropy
expressions containing $\mathfrak{N}$, such as \eqref{eq:klSI},
are of little value. Indeed, for a system containing $N$ (out of
$\mathfrak{N}$) distinguishable identical classical particles, the
entropy $S(N,\ldots)$, as a thermodynamic potential, must be replaced
by the reduced entropy
\begin{equation}
R(N,\ldots)=S(N,\ldots)-k\ln\frac{\mathfrak{N}!}{(\mathfrak{N}-N)!}\label{eq:ReducedEntropy}
\end{equation}
(defined in section~5.5 in \cite{Peters2010EJP}) which, unlike $S(N,\ldots)$,
does not depend on $\mathfrak{N}$ but instead happens to agree with
the entropy of indistinguishable particles (cf. section~5.6.6.2 in
\cite{Peters2010EJP}). This being said, $\mathfrak{N}$ is not necessarily
unknown and large compared to $N$; by exploiting some leeway in the
choice of what properties are regarded as permanent (cf. section~2.6.2
in \cite{Peters2010EJP}), it is, for example, possible to achieve
$\mathfrak{N}=2N$ in the above situation.

\section{Summary\label{sec:Summary}}

In section~\ref{sec:Einleitung}, a distinction was made between
the Gibbs paradox of the first kind (GP1) and that of the second
kind. The meanings of the terms \textit{identical} and \textit{indistinguishable}
were outlined. It was explained why the usual demonstration and resolution
of the GP1 are unsatisfactory. In section~\ref{sec:Demonstration-des-GP1},
an entropy decrease was calculated from the partitioning of an ideal
gas consisting of pairwise distinct bucky\-balls, each individually
made up of the isotopes carbon-12 and carbon-13. This specific
example demonstrates that the GP1 poses a clear contradiction to
the second law of thermodynamics and that not only classical but
also quantum systems can suffer from the GP1. In section~\ref{sec:Physikalischer-Ursprung-und},
it was shown that, after partitioning, the uncertainty about which
buckyballs are located in which subsystem, for one thing, increases
the entropies of the subsystems and, for another thing, causes a correlation
between the subsystems which deprives the entropy of its additivity.
These two effects compensate for the paradoxical entropy decrease
calculated in section~\ref{sec:Demonstration-des-GP1}. In section~\ref{sec:ClassicalResolution},
the concepts of section~\ref{sec:Physikalischer-Ursprung-und} were
used to resolve the GP1 in its original classical form.
\ack

The author thanks Alan Stern for valuable suggestions.

\appendix\section*{Appendix A}\setcounter{section}{1}\label{AppendixA}As
a general rule, one obtains the energy $E(T,\ldots)$ of a system
in canonical equilibrium from its partition function $Z(T,\ldots)$
by 
\begin{equation}
E(T,\ldots)=kT^{2}\frac{\partial}{\partial T}\ln Z(T,\ldots)\label{eq: E(T)}
\end{equation}
\cite{Feynman}. Applying \eqref{eq: E(T)} to \eqref{eq:ZiG(oBsT)}
yields $\mathbb{S}_{0}$'s energy 
\begin{eqnarray}
E_{0}(T,2V,2N) & = & 3NkT\:,\label{eq:E0(T,V,N)}
\end{eqnarray}
and applying it to \eqref{eq:ZustdTEils(fehlerhaft)} yields the same
energy for $\mathbb{S}_{1+2}$:
\begin{eqnarray}
E_{1+2}(T,2V,2N) & \:\underset{\hphantom{\eqref{eq:E0(T,V,N)}}}{=} & \: E_{1}(T,{\textstyle V},N)+E_{2}(T,{\textstyle V},N)\nonumber \\
 & \:\overset{\eqref{eq:ZustdTEils(fehlerhaft)}}{\underset{\hphantom{\eqref{eq:E0(T,V,N)}}}{=}} & \:2E_{0}(T,{\textstyle V},N)\nonumber \\
 & \:\overset{\eqref{eq:E0(T,V,N)}}{=} & \:3NkT\:.\label{eq:E1+2}
\end{eqnarray}
 
\appendix\section*{Appendix B}\setcounter{section}{2}\label{AppendixB}Let
$\mathsf{m}$ be a microstate of $\mathbb{S}_{1}$ and let $\left.\zeta\right.$
be $\mathbb{S}_{1}$'s particle composition in $\mathsf{m}$. The
conditional probability of $\mathsf{m}$ given that $\mathbb{S}_{1}$'s
particle composition is $\zeta$ is defined as
\begin{equation}
P\left(\mathsf{m}\!\mid\!\zeta\right)=\frac{P(\mathsf{m})}{P\left(\zeta\right)}\:,\label{eq:P(m)mitBedWkeit}
\end{equation}
where 
\begin{equation}
P(\zeta)=\underset{\textrm{of }\mathbb{S}_{1}\textrm{ is }\zeta}{\underset{\textrm{particle composition}}{\underset{\textrm{ of }\mathbb{S}_{1}\textrm{ in which the}}{\underset{\textrm{all microstates }\mathsf{n}}{\sum}}}}\; P(\mathsf{n})\label{eq:CompProb}
\end{equation}
is the probability that $\mathbb{S}_{1}$'s particle composition is
$\zeta$. The assumption that $\mathbb{S}_{1}$ contains $N$ particles
(cf. second remark of section~\ref{sec:Physikalischer-Ursprung-und})
guarantees 
\begin{equation}
\left.\zeta\in\{\zeta_{1},\,\zeta_{2},\,\ldots,\,\zeta_{{2N \choose N}}\}\right.\label{eq:ZetaAssump}
\end{equation}
and yields the normalization condition
\begin{equation}
\underset{i=1}{\overset{{2N \choose N}}{\sum}}P\left(\zeta_{i}\right)\overset{\eqref{eq:CompProb}}{=}\underset{\textrm{all microstates }\mathsf{n}\textrm{ of }\mathbb{S}_{1}}{\sum}\; P(\mathsf{n})\;=\;1\:.\label{eq:Normalization}
\end{equation}
For reasons of symmetry all possible particle compositions are equiprobable,
that is,
\begin{equation}
P(\zeta_{1})=P(\zeta_{2})=\ldots=P(\zeta_{{2N \choose N}})\overset{\eqref{eq:Normalization}}{=}\frac{1}{{2N \choose N}}\:.\label{eq:Harmonizitaet}
\end{equation}
(In the terminology of \cite{Peters2010EJP}, $\mathbb{S}_{1}$ is
called \textit{harmonic}.) Furthermore, in canonical equilibrium,
the conditional probability $P\left(\mathsf{m}\!\mid\!\zeta\right)$
is equal to $\mathsf{m}$'s probability in the hypothetical system
$\mathbb{S}_{1}^{\zeta}$. (This plausible claim is taken for granted
here; it can be verified by comparing the master equations underlying
$\mathbb{S}_{1}$ and $\mathbb{S}_{1}^{\zeta}$.) Expressing $\mathsf{m}$'s
probability in $\mathbb{S}_{1}^{\zeta}$ by the partition function
of $\mathbb{S}_{1}^{\zeta}$ yields 
\begin{equation}
P\left(\mathsf{m}\!\mid\!\zeta\right)=\frac{1}{Z_{1}\left(T,{\textstyle V},{\textstyle N}\mid\zeta\right)}\exp\left(-\frac{E_{\mathsf{m}}}{kT}\right)\:.\label{eq:bedWkeit}
\end{equation}
Note that using the partition function of $\mathbb{S}_{1}^{\zeta}$
is justified because, since there is no uncertainty about the particle
composition of $\mathbb{S}_{1}^{\zeta}$, $\mathbb{S}_{1}^{\zeta}$
does not suffer from $\mathbb{S}_{1}$'s non-ergodicity. Finally,
\begin{eqnarray}
P(\mathsf{m}) & \:\overset{\eqref{eq:P(m)mitBedWkeit}}{\underset{\hphantom{\eqref{eq:P(m)mitBedWkeit},\eqref{eq:Harmonizitaet},\eqref{eq:bedWkeit}}}{=}} & P\left(\zeta\right)\: P\left(\mathsf{m}\!\mid\!\zeta\right)\nonumber \\
 & \:\overset{\eqref{eq:ZetaAssump},\eqref{eq:Harmonizitaet},\eqref{eq:bedWkeit}}{=} & \frac{1}{{2N \choose N}}\frac{1}{Z_{1}\left(T,{\textstyle V},N\mid\zeta\right)}\exp\left(-\frac{E_{\mathsf{m}}}{kT}\right)\nonumber \\
 & \:\overset{\eqref{eq:ZetaAssump},\eqref{eq:Z1(T,V,Zeta_i)},\eqref{eq:ZustdTeils}}{\underset{\hphantom{\eqref{eq:P(m)mitBedWkeit},\eqref{eq:Harmonizitaet},\eqref{eq:bedWkeit}}}{=}} & \frac{1}{Z_{1}\left(T,{\textstyle V},N\right)}\exp\left(-\frac{E_{\mathsf{m}}}{kT}\right)\label{eq:P(m)exakt}
\end{eqnarray}
shows that $\mathbb{S}_{1}$ satisfies assumption \eqref{eq:P(m)/P(n)}
despite the fact that it is non-ergodic.

As a side note, if microstates of $\mathbb{S}_{1}$ with particle
numbers other than $N$ were not ignored (cf. second remark of section~\ref{sec:Physikalischer-Ursprung-und})
or if the ground-state energies of the buckyballs were not set to
the same value, then $\mathbb{S}_{1}$ would not satisfy assumption
\eqref{eq:P(m)/P(n)} and, as a consequence, using $\mathbb{S}_{1}$'s
partition function (e.g., to calculate $\mathbb{S}_{1}$'s entropy)
would lead to incorrect results.%

\bibliographystyle{iopart-num}
\bibliography{references}

\end{document}